\documentclass[twocolumn,showpacs,preprintnumbers,prb,amsmath,amssymb]{revtex4}

\usepackage{graphics}
\usepackage{amssymb}
\usepackage{amsmath}
\usepackage{overpic}
\usepackage{epstopdf}

\newcommand{\be}{\begin{equation}}
\newcommand{\ee}{\end{equation}}
\newcommand{\bea}{\begin{eqnarray}}
\newcommand{\eea}{\end{eqnarray}}

\newcommand{\ybco}{YBa$_{2}$Cu$_{3}$O$_{7-x}$}
\newcommand{\bscco}{Bi$_{2}$Sr$_{2}$CaCu$_{2}$O$_{8+x}$}

\newcommand{\BKT}{Berezinskii-Kosterlitz-Thouless}
\newcommand{\QCP}{quantum critical point}
\def\m{\mu}
\def\l{\lambda}
\def\th{\theta}

\def\s{\sigma}

\def\ra{\rightarrow}

\def\Ra{\Rightarrow}

\def\lb{\label}
\def\pref#1{(\ref{#1})}

\newcount\bozza \bozza=1
\ifnum\bozza=1
\newdimen\shift \shift=-2truecm
\def\lb#1{%
{\label{#1}\rlap{\kern\shift{$\scriptstyle#1$}}}}
\else\def\lb#1{\label{#1}} \fi

\begin{document}

\title{Robustness of the Berezinskii-Kosterlitz-Thouless Transition in Ultrathin NbN Films near the Superconductor-Insulator Transition}
\author{Jie Yong}
%%//\email{Present address: Center for Nanophysics and Advanced Materials, Department of Physics, University of Maryland, College Park, MD, 20742. Email: jyong@umd.edu}
\email{jyong@umd.edu}
\author{T. R. Lemberger}
\affiliation{Department of Physics, The Ohio State University, Columbus, OH, USA}

\author{L.~Benfatto} 
\affiliation{ISC-CNR and Dept. of Physics, Sapienza University of Rome,
  P.le A. Moro 5, 00185, Rome, Italy}

\author{K. Il’in, M. Siegel}
\affiliation{Institute of Micro- and Nano-electronic Systems, Karlsruhe Institute of Technology, Hertzstrasse 16, D-76187 Karlsruhe, Germany}

\date{\today}

\begin{abstract}
Occurrence of the \BKT ~(BKT) transition is investigated by superfluid density measurements for two-dimensional (2D) disordered NbN films with disorder level very close to a superconductor-insulator transition (SIT). Our data show a robust BKT transition even near this 2D disorder-tuned quantum critical point (QCP). This observation is in direct contrast with previous data on deeply underdoped quasi-2D cuprates near the SIT. As our NbN films approach the \QCP, the vortex core energy, an important energy scale in the BKT transition, scales with the superconducting gap, not with the superfluid density, as expected within the standard 2D-XY model description of BKT physics.
\end{abstract}

% fluctuations in sc, quantum critical phenomena, disorder effects in Tc, penetration depth

\maketitle

\section{Introduction}
\BKT  ~(BKT) transition, the only phase transition which can occur in the two-dimensional XY (2D-XY) model without breaking the continuous symmetry of the model,\cite{bkt} has generated great interest in condensed matter community for many years. It has been used to describe the superconductor-to-normal-metal thermal phase transition in 2D superconducting films in the context of free vortices emerging from a bath of thermally excited vortex-antivortex (V-aV) pairs, instead of breaking of Cooper pairs themselves.

There are several predicted experimental signatures of this transition.\cite{review_minnaghen} For example, above the transition temperature, the coherence length would diverge exponentially in the distance from the transition instead of the usual power-law, leading to a peculiar temperature dependence of the resistivity above the transition.\cite{nelson} But this temperature range is usually exceedingly small.\cite{nelson,hebard,mondal_bkt} The most direct and convincing evidence is that, universal and discontinuous drop in superfluid density is expected at the transition temperature.\cite{nelson_prl77} This has been shown beautifully in superfluid helium-4 system.\cite{helium4} In ultrathin conventional (Mo$_{77}$Ge$_{23}$,\cite{MoGe} InO$_x$\cite{armitage_prb07,armitage_prb11}, NbN\cite{mondal_bkt,kamlapure_apl10}) and quasi-2D cuprate (\ybco\cite{3dxy,hardy,yuri}, \bscco\cite{yong}) superconducting films, however, results are rather complex: (1) While significant drops in superfluid densities are indeed observed, they do not occur where 2D-XY model predicts - in ultrathin conventional films, they occur earlier than expected.\cite{MoGe, armitage_prb07, kamlapure_apl10,mondal_bkt} (2) In strongly underdoped cuprates, thick films of \ybco \cite{yuri} and \bscco,\cite{yong} and crystals of \ybco,\cite{hardy} thermal critical fluctuations are not observed near $T_c$, even though samples near optimal doping do exhibit critical fluctuations.\cite{3dxy,yong} It is worth noting that within the context of layered cuprates the possibility to identify BKT features associated to each bilayer unit relies on the general expectation that layers are weakly coupled. Thus, it is particularly surprising that when cuprate films are underdoped to near a superconductor-insulator transition (SIT), any thermal critical behavior, evidenced by the sharp downturn of superfluid density, disappears and the T-dependence of superfluid density goes quasi-linearly with the temperature all the way to T$_c$\cite{hardy,yuri,yong}. This contradicts the fact that underdoping usually increases anisotropy in cuprates, so that thermal critical behavior should be more robust than its counterpart near optimal doping. 

All the observations above indicate that there are several physical mechanisms at play in 2D superconducting films that are not captured by the 2D-XY model description of the BKT physics. First, as pointed out by one of the authors here, the relative energy scales involved in the BKT transition might not be universal after all\cite{benfatto_review}. For instance, the experimental data can be described well by allowing the ratio between the vortex core energy $\mu$ and the superfluid density to deviate from the 2D-XY model value. A smaller  (compared to the 2D-XY model prediction) or a larger $\mu$ can be used to fit the data of ultrathin conventional superconducting films\cite{kamlapure_apl10,mondal_bkt} and layered cuprates\cite{benfatto_bilayer,benfatto_mu} respectively, to account for the early or late drop in superfluid density.

Second, as evidenced mainly by scanning tunneling microscopy, intrinsic inhomogeneities emerge in these films of both conventional \cite{sacepe11,mondal_prl11,noat_cm12,lemarie_cm12} and cuprate\cite{gomes} superconductors especially when they are underdoped or driven to a very high disorder level. These inhomogeneities tend to broaden the transition and smear out the discontinuous drop. Any quantitative analysis then must take the local distribution of the superfluid densities into account\cite{benfatto_bilayer,benfatto_inho}. This will also complicate the analysis.

Third, when the system is pushed near the verge of a SIT, no matter by disorder or underdoping, there will be quantum fluctuations near such a quantum critical point (QCP). For example, there can be quantum V-aV pairs existing even at zero temperature.\cite{quantum_v} How quantum fluctuations affect BKT transition is still an open question. Goldman et al. suggested recently that macroscopic quantum tunneling in non-uniform thin films prevent resistivity dropping to zero below BKT transition.\cite{goldman} Previous studies on deeply underdoped cuprates also show that thermal critical behavior (BKT physics) persists in ultrathin films\cite{hetel} but it disappears in thick films near the QCP\cite{yuri,yong}.  This phenomenon has been shown to be universal in cuprates because it is robust against huge differences in anisotropy (\ybco vs \bscco) and disorder (thick films vs crystals). It seems BKT physics surrenders to quantum effects near a quantum critical point.

The main purpose of this article is to study the evolution of BKT physics near a QCP in 2D conventional superconducting films. While there are some previous superfluid density studies on 2D superconducting films, none of these films have been pushed to extremely high disorder so that a superconductor-to insulator (SIT) transition can be seen. In our case, by reducing the thickness and adding disorder in NbN films, we are able to drive NbN films with T$_c\sim15K$ smoothly all the way to insulating. We are able to push the disorder smoothly to SIT and still have reasonably sharp transitions. Temperature dependences of superfluid densities in these films are measured by a two-coil apparatus. Qualitatively, BKT transitions are observed for all the films even on the verge of SIT. By analyzing the data within the same theoretical scheme proposed in a previous article,\cite{mondal_bkt} we also observed an increase of both vortex-core energy and inhomogeneities as function of disorder, that consistently extend the previous studies on intermediately disordered films to highly disordered films near a 2D-QCP. This indicates that BKT physics remains robust against high disorder level or other quantum effects near a QCP in conventional 2D superconducting films. This robustness is in direct contrast with similar studies on layered underdoped cuprates, where BKT physics vanishes. We conclude this difference is because in deeply underdoped layered cuprates the increase of both the vortex-core energy and of the coherence length can conspire to mask the occurrence of 2D behavior near the SIT.

\section{Experimental}
The superconducting NbN films were deposited by reactive magnetron sputtering of a pure Nb target in an Ar+N$_2$ gas mixture, at a total pressure of about $10^{-3}$ mbar. The epipolished R-plane sapphire substrates were kept at 550$^\circ$C during the film growth. Deposition rate (0.17nm/sec) is calibrated so thickness of the film is inferred from the sputtering time. The deposition process was optimized with respect to the partial pressure of N$_2$ and the deposition rate to provide the highest transition temperature for films with the smallest studied thickness. More details of growth can be found in this paper\cite{growth}. Many physical parameters have been measured for these NbN films.\cite{para} They are patterned to nanowires and used to make single photon detectors.\cite{growth} We emphasize that the films are homogeneously disordered because (1) Sheet resistance is almost a constant above the transition temperature and does not show any discontinuity at higher temperature. (2) Conducting films with reasonably sharp and single transition can be grown with the thickness of only three or four unit cell. (3) It is generally easier to be homogeneously disordered for a binary compound, like well-studied InO$_x$ and TiN films.

\begin{figure}[!t] \centering
  \resizebox{8cm}{!}{
  \includegraphics[width=8cm]{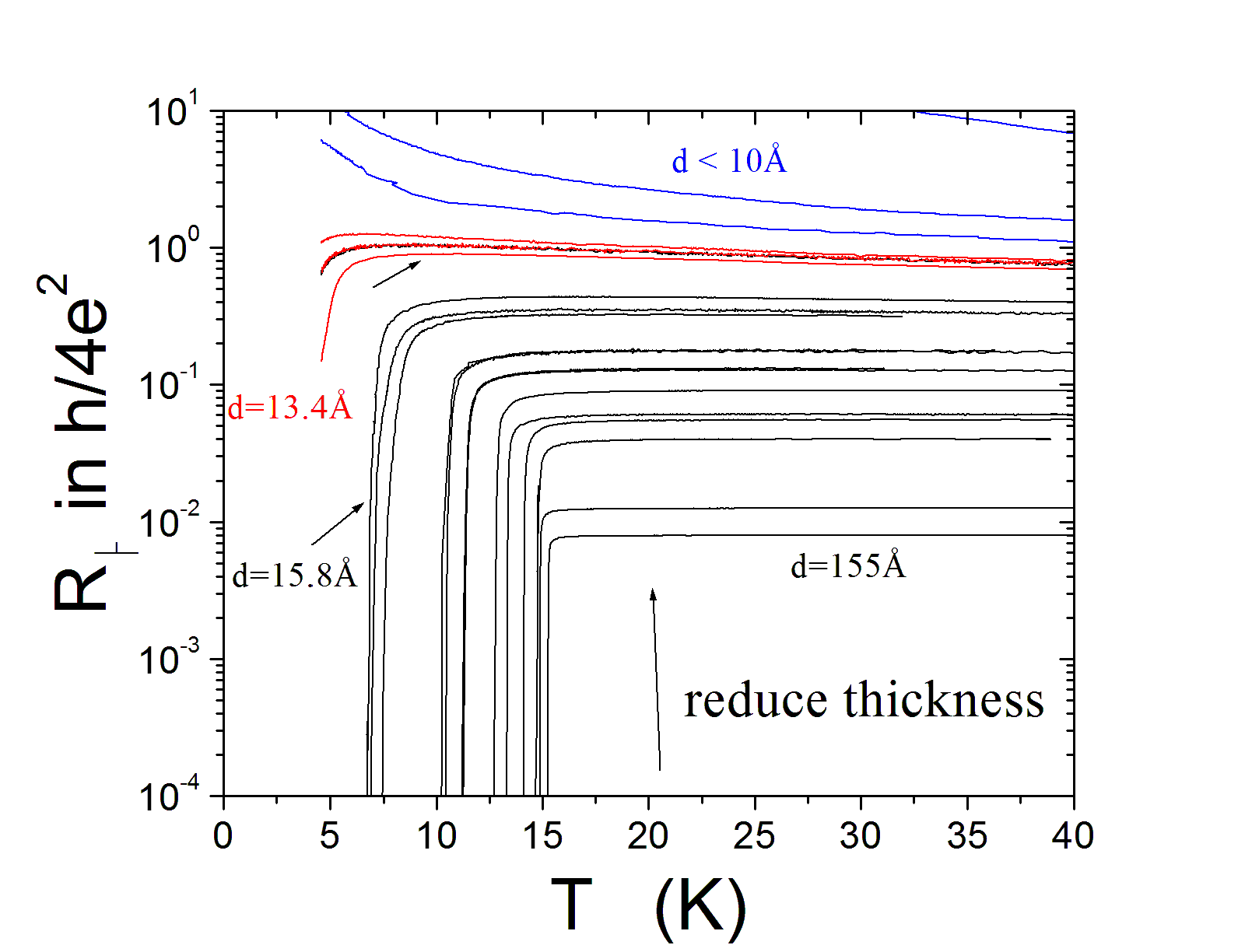}}
\caption{Sheet resistance $R$, normalized by quantum resistance $h/4e^2$ , vs. temperature for many NbN films with different thickness $d$ and disorder. Superconductor-insulator transition is observed for films with $R(20 K) < h/4e^2$ (black curves). Some films (red curves) are very close to SIT but the measurement temperatures go down only to 4.2K.}
\label{fig1}
\end{figure}

Figure \ref{fig1} shows that a nice superconductor-insulator transition is observed when the film thickness is reduced to a few unit cell\cite{unitcell} (one unit cell = 0.44nm). Superconducting and insulating films are separated by quantum resistance for cooper pairs, which is $h/4e^2$ = 6.45 k$\Omega$. Thick NbN films have $T_c \approx 15 K$, close to the bulk value. As films get thinner or more disordered, sheet resistance increases and $T_c$ drops. Near SIT, we are able to consistently reproduce films with $T_c$ about 4K with sheet resistance $\sim 5.5k\Omega$ above T$_c$. Unfortunately, these ultrathin films degrade in the air and the effective sheet resistance will change. Both the superfluid density and $T_c$ will change too. This degradation prevents us from directly comparing resistance and superfluid data on the same film. But it gives us another way to tune the SIT without sample-to-sample variation. As we will show later, in terms of SIT tuning, there is no difference among thickness, disorder or degradation.

Superfluid densities are measured by a two-coil mutual inductance apparatus.\cite{measure} The film is sandwiched between two coils, and the mutual inductance between these two coils is measured at a frequency $\omega/2\pi$ = 50 kHz. The measurement actually determines the sheet conductivity, $Y \equiv (\sigma_1 + i\sigma_2)d$, with $d$ being the superconducting film thickness and  $\sigma$ being the conductivity. Given a measured film thickness, $\sigma$ is calculated as: $\sigma = Y/d$. The imaginary part, $\sigma_2$, yields the superfluid density through: $\omega\sigma_2\equiv n_se^2/m$, which is proportional to the inverse penetration depth squared: $\lambda^{-2}(T) \equiv \mu_0\omega\sigma_2(T)$, where $\mu_0$ is the permeability of vacuum. As is customary, we refer to $\lambda^{-2}$ as the superfluid density. The dissipative part of the conductivity, $\sigma_1(T)$, has a peak near T$_c$, whose width provides an upper limit on the spatial inhomogeneity of T$_c$ over the 10 $mm^2$ area probed by the measurement. Data are taken continuously as the sample slowly warms up so as to yield the hard-to-measure absolute value of $\lambda^{-2}$ and its T-dependence. This two-coil technique is powerful \cite{benfatto_review} to study thermal critical behavior like BKT transition near T$_c$. It is also unique for 2-D films because it can give sheet superfluid density $d/\l^2(T)$ without the knowledge of the film thickness, which is the case here.

\begin{figure}[!t] \centering
  \resizebox{8cm}{!}{
  \includegraphics[width=8cm]{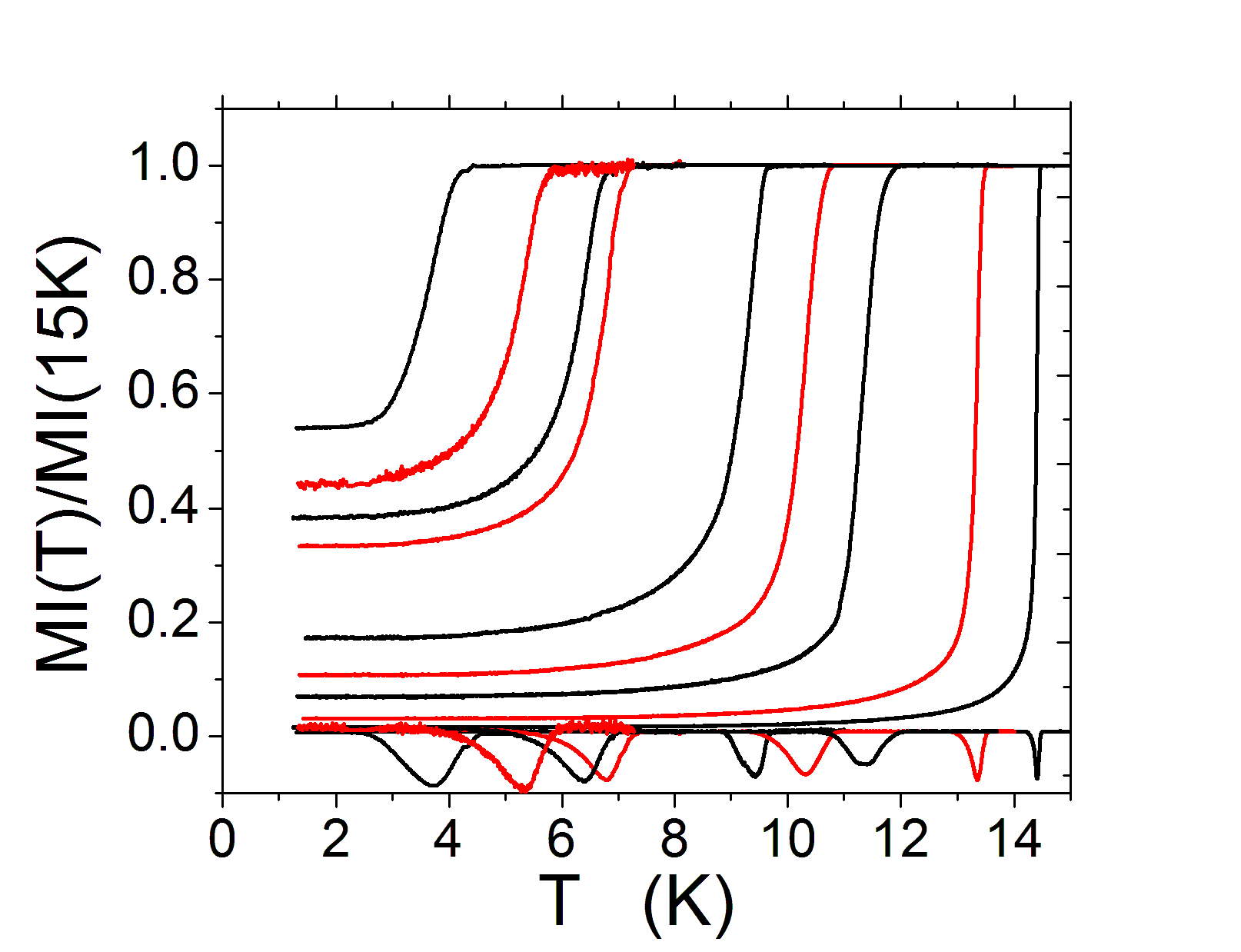}}
\caption{Temperature dependence of mutual inductance, MI(T)/MI(15 K), including both real (upper curves) and imaginary parts (lower dips), of many NbN films with different thickness and disorder level. For clarity purposes, not all data are shown.}
\label{fig2}
\end{figure}

Mutual inductance data, MI(T)/MI(15 K), of some films are shown in Figure \ref{fig2}. Nice sharp transitions are observed for most films. As films get thinner, or get more disordered, or simply degraded, $T_c$ drops and the normalized mutual inductance at $T \ll T_c$ grows. This shows that the ability of film to screen magnetic field gets reduced as the film get thinner. This ability is directly related to the areal superfluid density of the film. The imaginary part of the mutual inductance, which shows a dip at the transition, relates to the dissipation of moving vortices. The width of the dip shows that inhomogeneity gets larger as the film disorder is increased to near SIT, consistent with other local gap measurements.\cite{mondal_prl11}

\section{Theoretical analysis of the data}
\label{theo}

We start by briefly reviewing the basic elements of the BKT transition that will be needed for the fitting of the experimental data.\cite{benfatto_review} As we mentioned at the beginning, the BKT transition was originally formulated within the context of the two-dimensional (2D) $XY$-model, which describes the exchange
interaction between classical two-component spins with fixed length $S=1$: 
\be
%\lb{xy}
H_{XY}=-J\sum_{\langle ij \rangle}\cos (\th_i-\th_j), \label{xy}
\ee
where $J$ is the spin-spin coupling constant and $\theta_i$ is the angle that the $i$-th spin forms with a given direction, and $i$ are
the sites of a square lattice. Within the context of 2D superconductors $\theta_i$ plays the role of the SC phase, and J (now written as $J_s$, and referred to as the superfluid stiffness) is connected to the areal density of superfluid electrons $\rho_s^{2d}\equiv n_s d$, which in turn is measured via the inverse penetration depth $\lambda$ of the magnetic field:
\be
\label{defj}
J_s =\frac{\hbar^2\rho_s^{2d}}{4m}=\frac{\hbar^2 d}{4 e^2 \mu_0 \l^2 }.
\ee
Usually both quasiparticle excitations and phase fluctuations contribute to the depletion of $J_s$ towards zero. In the case of our NbN films the quasiparticle contribution can be well accounted by the dirty-limit BCS expression,
\be
\label{jbcs}
\frac{J^{BCS}(T)}{J^{BCS}(0)}=\frac{\Delta(T)}{\Delta(0)}\tanh
\left[\frac{\Delta(T)}{2k_B T} \right],
\ee
by using eventually $\Delta(0)/T_{BCS}$ as a free parameter, to account for the relatively large $\Delta(0)/T_c$ ratio reported in NbN as
disorder increases.\cite{mondal_prl11}. 

For what concerns transverse (i.e. vortical) phase fluctuations their effect will be accounted for by numerical solution of the BKT renormalization-group (RG) equations, whose relevant variables are the dimensionless quantities\cite{bkt,review_minnaghen,benfatto_review}:
\bea
\label{defk}
K(0)&=&\frac{\pi J^{BCS}(T)}{T},\\
\label{defg}
g(0)&=&2\pi e^{-\beta\mu},
\eea
where $\mu$ is the free energy of a vortex core, with radius about equal to the superconducting coherence length, $\xi(T)$, and $g$ is called the vortex fugacity. Notice
that $J^{BCS}$ enters here to determine the initial value of $K$, i.e. its short-distance value. Its long distance value follows 
by the solution of the well-known RG equations\cite{bkt,review_minnaghen,benfatto_review}:
\bea
\label{eqk} 
\frac{dK}{d\ell}&=&-K^2g^2,\\ 
\label{eqg} 
\frac{dg}{d\ell}&=&(2-K)g.
\eea
where $\ell \equiv \ln (r/\xi)$ is the rescaled length scale. The observed superfluid
density is identified by the limiting value of $K$ as one goes to large distances\cite{nelson_prl77}:
\be
\label{jlim}
J_s\equiv \frac{T K(\ell\ra \infty)}{\pi}.
\ee

\begin{figure}[!t] \centering
  \resizebox{8cm}{!}{
  \includegraphics[width=8cm]{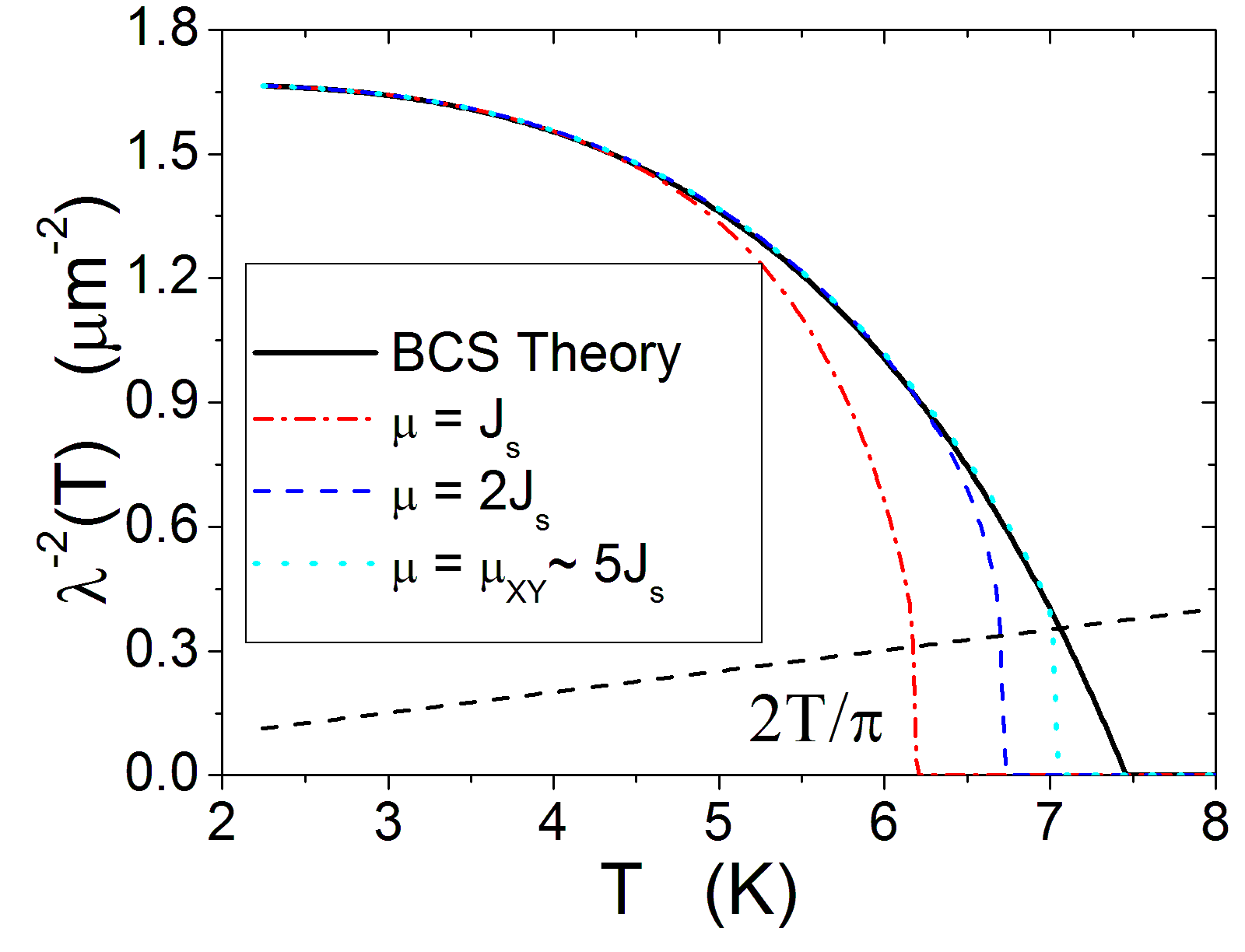}}
\caption{Role of the vortex-core energy on the BKT transition. The solid black line represents the temperature dependence of $1/\l^2$ within BCS theory, as described by Eq.\ \pref{jbcs}, for typical parameter values appropriate for NbN films. The BKT transition temperature depends on the value of the vortex-core energy. For $\mu$ as large as in the $XY$ model the transition occurs (dotted green line) at the intersection between the BCS curve and the universal line $2T/\pi$. However, for smaller $\mu$ values, $J_s$ is renormalized with respect to its BCS counterpart already before the transition, so that the transition occurs at a lower $T_{BKT}$ (see dashed blue line and red dot-dashed line). Notice that in all these cases the universal relation \pref{jump} is satisfied, and $1/\l^2$ jumps discontinuously to zero after intersection with the $2T/\pi$ line. }
\label{fig_mu}
\end{figure}

The basic idea of the RG equations is to look at the large-scale behavior of the superfluid stiffness and of the vortex fugacity. When $g\ra 0$ it means that single-vortex excitations are ruled out from the system, which is then SC: indeed, as one can see from Eqs.\ \pref{eqk}-\pref{eqg} when $g\ra 0$, $K$ goes to a constant and then $J_s$ from Eq.\ \pref{jlim} is finite. If instead $g\ra \infty$ at large distances it means that vortices proliferate and drive the transition to the non-SC state, since $K\ra 0$. The large-scale behavior depends on the initial values of the coupling constants $K,g$, which in turn depend on the temperature. The BKT transition temperature is defined as the highest value of $T$ such that $K$ flows to a finite value, so that $J_s$ is finite. This occurs at the fixed point $K=2, g=0$, so that at the transition one always has:
\begin{equation}
\label{jump}
K(\ell\ra\infty,T_{BKT})=2, \Ra \frac{\pi J_s(T_{BKT})}{T_{BKT}} = 2,
\end{equation}
while above it, $J_s=0$. As a consequence, at $T_{BKT}$, $J_s$ jumps discontinuously from the universal value $2T_{BKT}/\pi$ to zero. However, it should be emphasized that already before $T_{BKT}$ the effect of short length-scale vortex-antivortex pairs is in general to deplete $J_s$ with respect to its initial value, given by
the BCS estimate \pref{jbcs}. This effect is usually negligible when $\mu$ is large, as it is the case for superfluid films\cite{helium4} or within the standard $XY$ model\cite{benfatto_review}, where $\mu_{XY}\sim (\pi^2/2) J_s$. In this case one can safely estimate $T_{BKT}$ as the temperature where the line $2T/\pi$ intersects the $J^{BCS}(T)$ from Eq.\ \pref{jbcs}, see Fig.\ \ref{fig_mu}. However, as $\mu$ decreases the renormalization of $J_s$ due to bound vortex pairs increases, and consequently the deviation of $J_s$ from its BCS counterpart starts considerably before the transition temperature itself,\cite{benfatto_mu,benfatto_review} see Fig. \ref{fig_mu}.

\begin{figure}[!t] \centering
  \resizebox{8cm}{!}{
  \includegraphics[width=8cm]{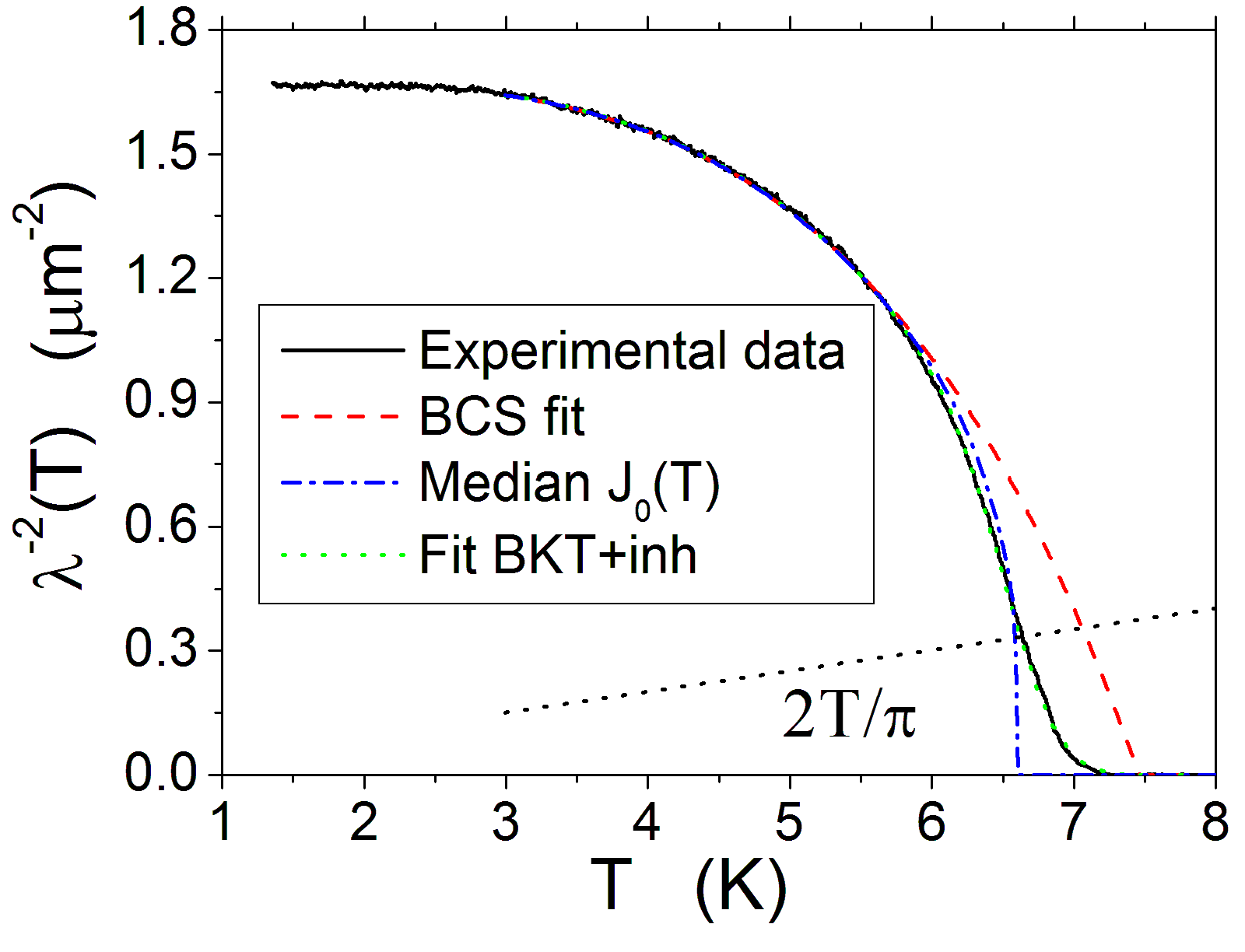}}
\caption{Role of inhomogeneity on the BKT transition. The experimental data (black curve) correspond to the sample labeled as S193v1 in Table I. While the median
 $J_0(T)$ (blue dot-dashed line) of the Gaussian distribution \pref{gauss} of possible $J_i$ realizations has a sharp transition, the average stiffness $J_{av}(T)$ from Eq.\ \pref{jav}
 vanishes with a smoother tail. The deviation from the BCS curve (red dashed line) before than the transition is due instead to the low value of the vortex-core energy, see Fig. \ \pref{fig_mu}.}
\label{fig_s89}
\end{figure}

At intermediate and strong disorder STS experiments have shown that NbN films\cite{mondal_prl11,noat_cm12,lemarie_cm12} exhibit
a spatial inhomogeneity of the SC spectra, that becomes particularly pronounced near the SIT. Even though STS spectra probe only the local
DOS of the sample, one would expect that the same inhomogeneity reflects also in the local superfluid stiffness. As a consequence, one
can imagine that the sample admits a given distribution of local $J_i$ values with probability density $P(J_i)$ and local BCS and BKT transition
temperatures $T_c^i$ and $T_{BKT}^i$, respectively. A possible phenomenological way to estimate the overall superfluid stiffness \cite{benfatto_inho} is to
compute the average $J_{av}$ as:
\be
\label{jav}
J_{av}(T)= \sum_i P(J_i) J^i_s(T),
\ee
where $P(J_i)$ can be taken for example as a Gaussian distribution centered around the experimental value of $J_0$ at $T=0$,
\be
\label{gauss}
P(J_i)=\frac{1}{\sqrt{2\pi}\s}
\exp\left[-(J_i-J_0)^2/2\s^2\right].
\ee
When $J_i=J_0$ the corresponding $J_s^i(T)\equiv J_0(T)$ coincides with the BKT curve obtained from the BCS fit \pref{jbcs} of the
experimental data, shown with a dot-dashed line in Fig.\ \ref{fig_s89}. For the remaining $J_i$ values we rescale the corresponding BCS temperatures as
$J_i/T_{BCS}^i=J_{BCS}(0)/T_{BCS}$ and we compute $J_s^i(T)$  and the corresponding BKT temperature $T_{BKT}^i$ by the numerical solution of the RG equations \pref{defk}-\pref{defg} above. Once obtained this set of $J_s^i(T)$ curves we compute at each temperature the average value $J_{av}(T)$ according to Eq.\ \pref{jav}. 
When all the stiffness $J_s^i(T)$ are different from zero, as is the case at low temperatures, the average stiffness will be centered around the center of the
Gaussian distribution \pref{gauss}, so that it will coincide with $J_0(T)$. However, by approaching $T_{BKT}$ defined by the average $J_0(T)$ not all the patches make the transition at the same temperature, so that the BKT jump is rounded and $J_{av}$ remains finite above the average $T_{BKT}$, in agreement with
the experiments, see Fig. \ref{fig_s89}. We note that Eq.\ \pref{jav} implies an average of the imaginary part of the complex optical conductivity, since $J_s\propto \l^{-2}\propto \sigma_2$. The same mechanism applied to its real part leads to a broadening of the $\sigma_1$ peak at the transition, as discussed in Refs.\ [\onlinecite{benfatto_bilayer,benfatto_review}]. Thus, one expects that a smearing of the abrupt superfluid-density jump due to increased inhomogeneity is also accompanied by  a broadening of the $\sigma_1$ peak, as we will discuss indeed in the next section in connection to the experimental data.

\section{data and discussion}
\begin{figure}[htb]
\begin{center}
\includegraphics[scale=0.3,clip=true]{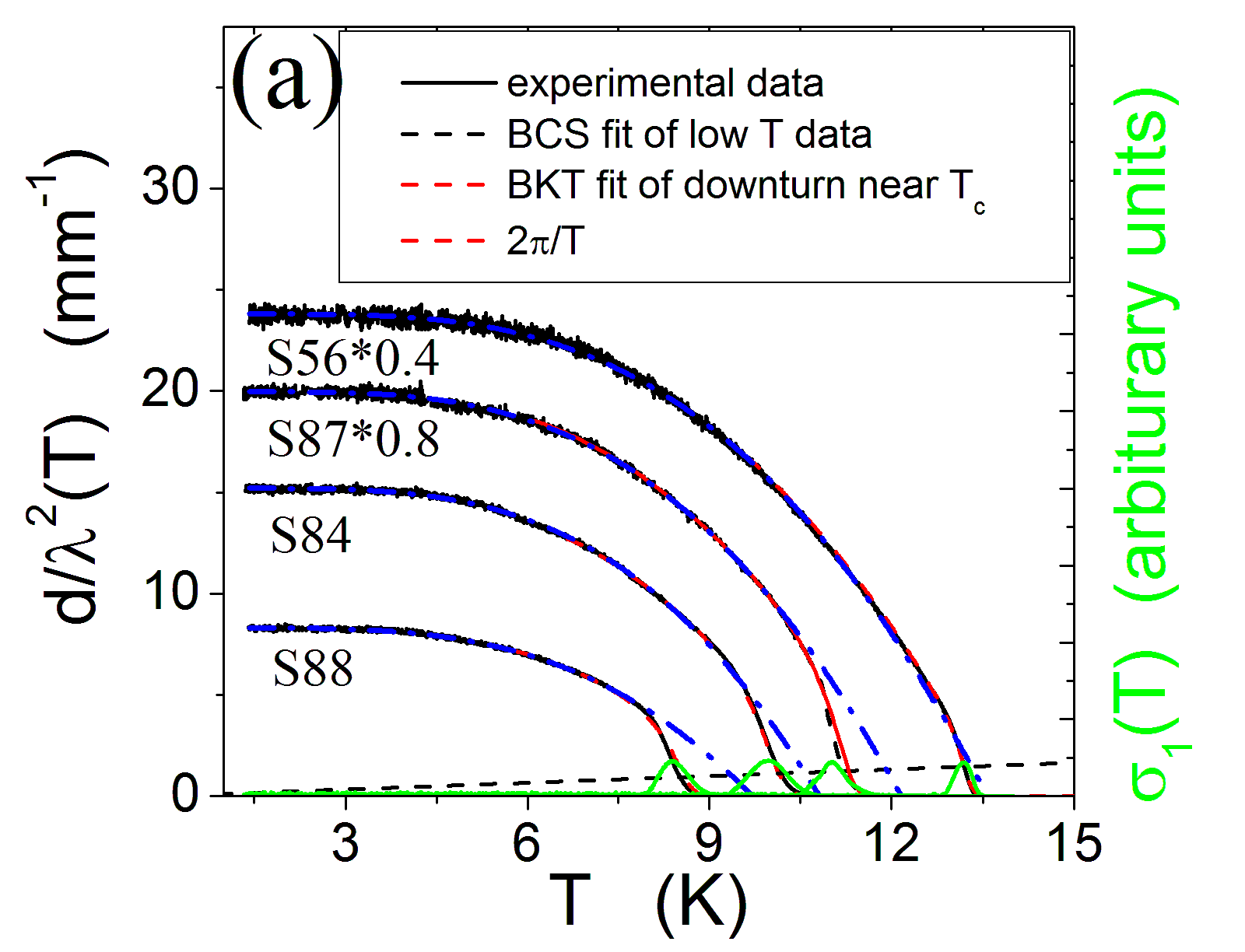}
\includegraphics[scale=0.3,clip=true]{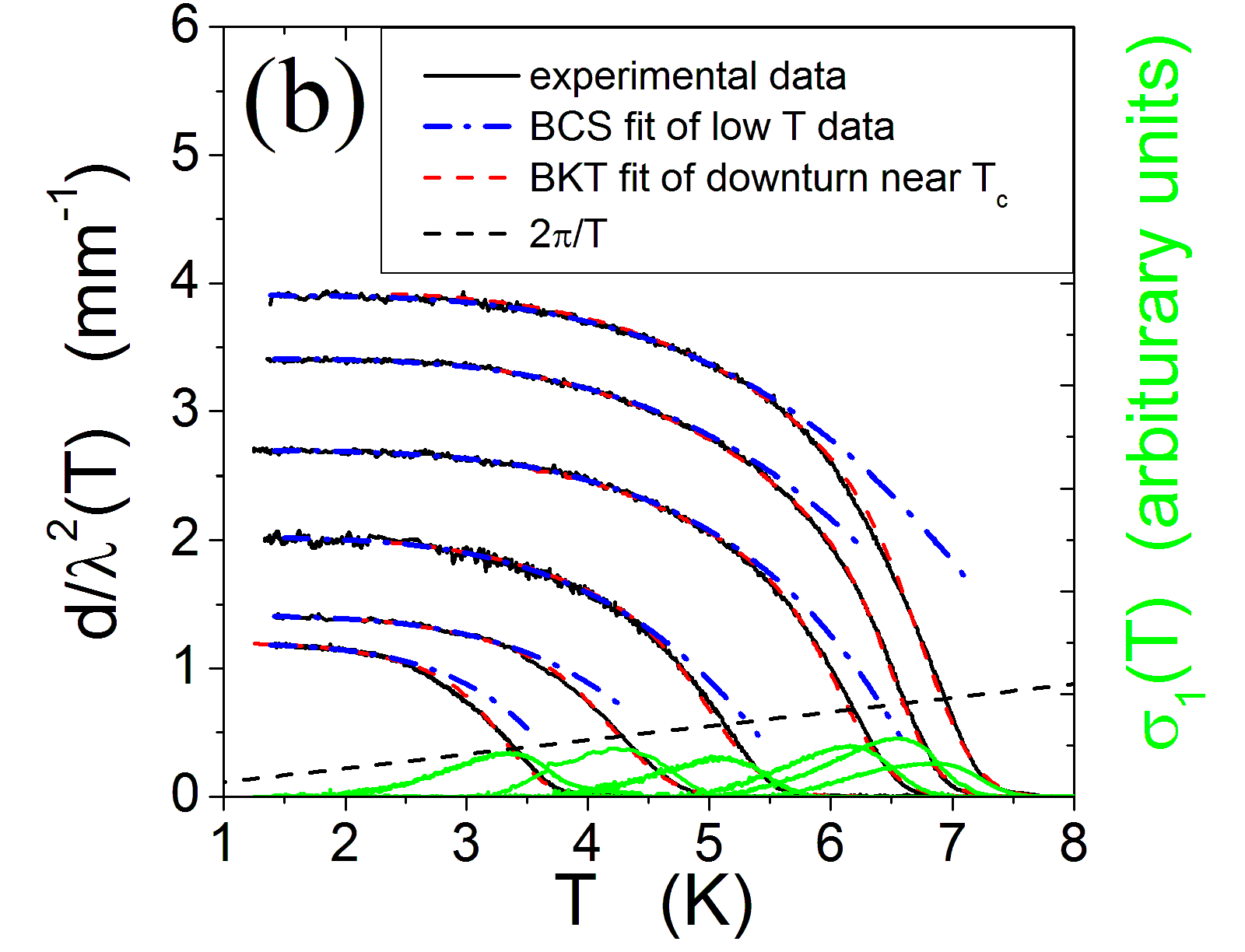}
\caption{Experimental data (solid black curves) of various samples [(a) for moderately disordered films and (b) for highly disordered films] are fitted by BCS dirty limit theory well (blue dash-dotted curves) until deviations occur. Black dashed line gives the prediction on where BKT transition should occur given by 2D XY model. The deviations are fitted by the procedure mentioned in Sec. III (red dashed curves). The fitting parameters values are reported in Table I.}
\label{fit_all}
\end{center}
\end{figure}

\begin{table}[t]
\begin{center}
\caption{Experimental values of nominal thickness $d$ and sheet superfluid density $d/\l^{2}(0)$, which are transferred to energy scale $J_s(0)$ via Eq. \pref{defj}, along with the best fit parameters.  Here the BCS transition temperature $T_{BCS}$ and the superconducting gap $\Delta(0)$ are obtained from the BCS fit. The vortex-core energy $\mu$ and the degree
of inhomogeneity $\delta$, both normalized by $J_s(0)$, are from best BKT fit. The temperature $T_{BKT}$ corresponds to the transition temperature of the
median $J_0(T)$ of the Gaussian distribution, see also Fig.\ \pref{fig_s89}.}

\label{t-table}
\begin{tabular}{|c|c|c|c|c|c|c|c|c|}
\hline 
Film ID & $d$&$d/\l^{2}(0)$ &  $J_s(0)$ &$T_{BCS}$ & $T_{BKT}$  &
$\m/J_s$ & $\Delta/J_s$ & $\delta/J_s$   \\
& $(nm)$&$(mm^{-1})$ & (K) &(K) & (K) & & \\
\hline
S56 & 5.5&59.4 & 368.28 &  13.7 & 13.23 & 0.6 & 0.08 & 0.007\\
S87 & 3.2&24.96 & 154.75 & 12.23 & 11.28 & 0.54 & 0.17 & 0.013 \\
S84 & 2.72&15.2 & 94.27 &11.1 & 10.1 & 0.7 & 0.253 & 0.02\\
S88 & 2.24&8.31 & 51.52 & 9.7  & 8.59 & 0.83 & 0.461 & 0.02 \\
S194 & 2.28&5.68 & 35.2 & 9.3 & 8.23 & 0.75 & 0.739 & 0.04 \\
S196v2&2.16& 4.02 & 	24.9	& 7.7	& 7.19 	& 1.65 &0.897	& 0.05 \\
S197   &2.1  	& 3.93	& 24.34	& 7.95	 & 6.9  	& 1.25&0.947	& 0.045 \\
S196v1	&2.16&3.82	& 23.69	& 8.68	& 7.45	& 1.55	&1.063& 0.045\\
S193v1	&2.04& 3.41	& 21.12	& 7.48	& 6.94	& 1.7	& 1.027& 0.042\\
S193v4	&2.04& 2.69	& 16.7	& 7.1	& 6.53	& 1.9	& 1.233&0.055\\
S193v5	&2.04& 2.58	& 16	& 6.78	& 6.24	& 1.93	&1.229& 0.062\\
S193v2	&2.04& 2.04	& 12.65	& 5.87	& 5.42	& 2.2	&1.346&0.07\\
S193v3	&2.04& 1.41	& 8.72	& 5.15	& 4.64	& 2.25&1.713	& 0.085\\
S89	        &1.58  & 1.20	& 7.41	&4.2	& 3.38& 	1.8	&1.474& 0.08\\

\hline 
\end{tabular}
\end{center}
\end{table}

The sheet superfluid densities $d/\l^{2}(T)$ of many films with different thicknesses and disorder are shown in Fig. \ref{fit_all}. Fig. 5(a) shows moderately disordered films with $T_c >8K$. Fig. 5(b) are films with $T_c$ less than 8K. The modified BKT theory developed in Sec. III, with vortex core energy a free parameter and consideration of inhomogeneities, fit the experimental data pretty well. Table I shows major fitting parameters for every sample studied here. Two energy scales, the vortex core energy $\mu$ and the SC gap $\Delta (0)$, normalized to the superfluid stiffness, are found to be correlated at high disorder.  This is shown in Fig. 6. The sheet superfluid density $d\l^{-2}(T)$ is converted to the superfluid stiffness $J_s$ by means of Eq.\ \pref{defj}. By using $\hbar^2/4e^2\mu_0 k_B=6.2 \times 10^{-3}$ Km we can express $J_s$ in K as:
\be
\label{defjk}
J_s[K] = 0.62 \, \frac{d[\AA]}{\lambda^2[\m m^2]}
\ee

\vspace{0.5cm}

Qualitatively, all films show a deviation from BCS theory fit earlier than what 2D XY model predicts, which is the intersection with the dashed line in Fig. \ref{fit_all}.  This means that the early appearance of the BKT downturn, first shown in moderately disordered Nb\cite{Nb} and NbN\cite{mondal_bkt} films, is a common characteristic of conventional 2D SC films.  As we discussed in Sec. III, a small value of the vortex core energy is responsible for the fact that $1/\lambda^2(T_{BKT})<1/\lambda^2_{BCS}(T_{BKT})$, so that also the $T_{BKT}$ in the perfectly homogeneous case would occur before than the 2D-XY model prediction. However, the presence of inhomogeneity smears out the sharp BKT drop and gives a finite width to the transition, as evidenced by small peaks in $\sigma_1$ near $T_c$. These peaks get wider as disorder increases, consistent with the increase in inhomogeneity observed in tunneling \cite{mondal_prl11,noat_cm12, lemarie_cm12}, and also with the increase of the superfluid-stiffness distribution width $\delta/J_0$ obtained by the BKT fit (see Table I).

Quantitative results are shown in Table I. We are able to tune the disorder so that the superfluid stiffness changes by a factor of 50 and $T_c$ changes by a factor of 3. Our most disordered film has a $T_c \sim 4K$ and $J_s(0) \sim 7.4K$ compared to  $T_c \sim 8K$ and $J_s(0) \sim 60K$ for the most disordered film in a previous study.\cite{mondal_bkt} $T_{BKT}$ is very close to $T_c$ at low disorder (3\% difference) and becomes more separated as disorder is increased. For the most disordered film S89, this difference is as large as 20\%, showing the separation of two energy scales $T_{BKT}$ and $T_c$. Interesting, the energy scale for sheet superfluid density $J_s(0)$ is getting close to the scale of $T_c$ and may become the limiting factor for $T_c$. Three fitting parameters, $\mu$, $\Delta(0)$, $\delta$, normalized by $J_s(0)$ are listed in Table I.

\begin{figure}[htb]
\begin{center}
\includegraphics[scale=0.3,clip=true]{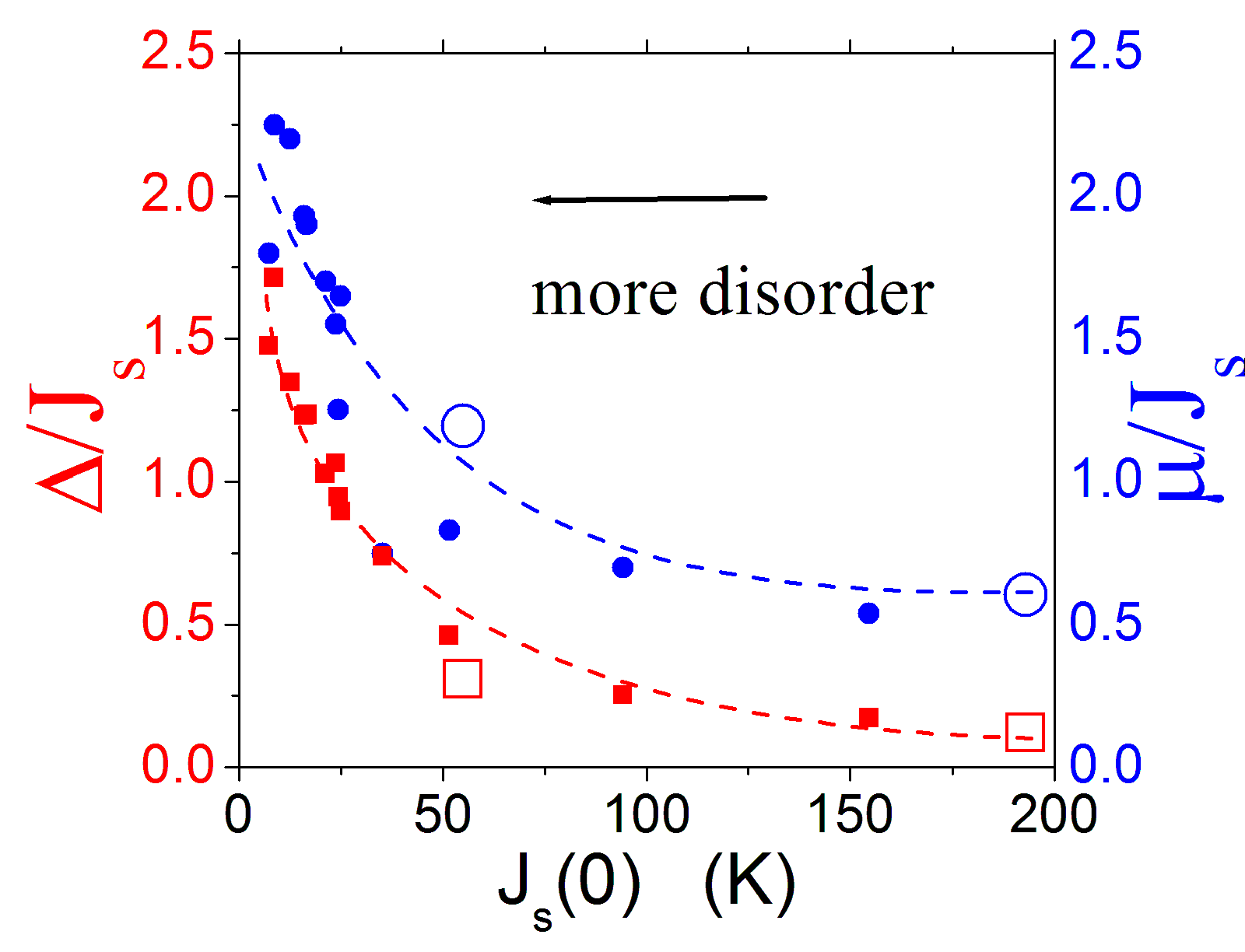}
\caption{Evolution of the vortex-core energy (blue circles) and SC gap (red squares), normalized by the superfluid stiffness $J_s(0)$, with $J_s(0)$ in our NbN films. Open symbols are data from Mondal et al.\cite{mondal_bkt} Dashed lines are guides to the eye.}
\label{fig_mudelta}
\end{center}
\end{figure}
One of our major findings is shown in Fig. \ref{fig_mudelta}. With $J_s(0)$ characterizing the disorder level of the films, we found that the two energy ratios $\Delta(0)/J_s(0)$ and $\mu/J_s(0)$ are highly correlated at high disorder. This means that near the quantum critical point, the vortex-core energy $\mu$, an important energy scale in BKT transition, does not scale with the superfluid stiffness, as given by 2D XY model. Instead, it scales with the superconducting gap, which is the pairing strength of the Cooper pairs.  

There are two different contexts our results can be put into. First, the observation of robust BKT transition is consistent with previous results on moderately disordered films\cite{mondal_bkt} and extends it to highly disordered films on the verge of the 2D superconductor-insulator transition (SIT). The vortex core energy is also shown to scale with the energy gap near the QCP. Our data on more than ten films firmly confirmed the observation of a previous study on three films, that can be understood theoretically\cite{mondal_bkt} in terms of the increasing separation between the energy scales associated to pairing and phase coherence induced by disorder. 

Second, the robustness of BKT transition observed here is in direct contrast with similar superfluid density studies on deeply underdoped {\em layered} cuprates\cite{hardy, yuri,yong} near the QCP, where no downturns are observed at all. In these deeply underdoped quasi-2D cuprates, superfluid density goes linearly with the temperature in almost all the compounds studied, including both \ybco ~and \bscco, both crystals and films. 

What is the big difference between the two systems? First of all, we should recall that the possibility to observe a BKT transition in thick cuprate films relies on the interplay between different length and energy scales. In cuprates, there are three length scales in the c-axis: the thickness $d_{film}$, the neighboring superconducting $CuO_2$ bilayer distance $d_{CuO_2}$ and the c-axis coherence length $\xi_c$. $d_{film}$ is typically several hundred nanometers for "thick" films and much larger for crystals. The distance between neighboring $CuO_2$ bilayers is about 12 $\AA$ in YBCO and 15 $\AA$ in Bi-2212. Only the coherence length $\xi_c$ has a temperature or doping dependence. Near well-studied optimal doping, $\xi_c \sim 2\AA$ , which is much less than $d_{CuO_2}$. We have:
\be
\label{2Dc}
\xi_c < d_{CuO_2} < d_{film}   ~~~(quasi-2D),
\ee
That is why cuprates are generally considered to be quasi-2D, and one would generically expect a BKT transition for each {\em isolated} layer. This means that the temperature where the universal jump Eq. \pref{jump} should occur is compared to the superfluid stiffness of a single bilayer, i.e. $d$ is replaced by $d_{CuO_2}$ in Eq.\ \pref{defj}. Nonetheless, layers are not completely independent, since the phase in neighboring layers is coupled by a (weak) Josephson-like coupling $J_\perp$. Once more, when $J_\perp$ is much smaller than in-plane stiffness one would expect BKT-like behavior. However, this is only true when the vortex-core energy is of the value expected in the XY model: indeed, it has been shown\cite{benfatto_mu} that for {\em larger} values of the vortex-core energy $\mu$, the BKT transition can occur at temperatures larger than expected within the XY model, since interlayer coupling predominates over vortex unbinding on a wider range of temperatures. The observation of sharp BKT downturns in optimally-doped \bscco\cite{yong} suggests that in this material not only is anisotropy very large (i.e. $J_\perp$ is very small), but also the vortex-core energy must be of the order of the XY-model value, allowing for a BKT transition controlled by the stiffness of each isolated bilayer. 

When doping is lowered both $\xi_c$ and $\mu$ might grow. From one side, if $\xi_c$ remains small, an increase of the ratio $\mu/J_s$ analogous to the one reported above for conventional films could by itself move the $T_{BKT}$ to higher temperatures, making also the jump barely visible. Such an increase of $\mu/J_s$ has been indeed inferred by a theoretical analysis similar to the one discussed above in ultra-thin \ybco films.\cite{benfatto_inho} In the case of conventional superconducting films the increase of $\mu/J_s(0)$ can be understood as an effect of the increasing separation between the pairing energy scale and the phase coherence due to disorder, which can be also be responsible for the pseudogap observed by STM in this material.\cite{sacepe11,mondal_prl11} A similar analysis in the context of cuprates would be very interesting, since it could shed new light on the effect of disorder on the underdoped regime of these materials as well.

On the other hand, if $\xi_c$ exceeds the $d_{CuO_2}$ and stays less than $d_{film}$, 
\be
\label{3Dc}
d_{CuO_2}<\xi_{c}<d_{film}   ~~~(3D),
\ee
then cuprates become more three dimensional. A BKT paradigm then does not apply and the characteristic BKT jump disappears\cite{nota_charged}. Eventually, when $\xi_c$ exceeds even the film thickness 2D behavior could be recovered again, but now $T_{BKT}$ should correspond to the sheet superfluid density of the whole film, which is very large for thick films. Thus, the BKT jump would become practically indistinguishable from the transition temperature due to other thermal excitations (as quasiparticle or longitudinal phase fluctuations). The in-plane coherence length $\xi$ is related to upper critical field $H_{c2}$ by:
\be
\label{coherence_length}
H_{c2}=\Phi_0/2\pi\xi^2,
\ee
where $\Phi_0$ is flux quantum. Therefore a large coherence length corresponds to a relatively small $H_{c2}$ for deeply underdoped cuprates. In this view, the lack of BKT signatures in superfluid-density data can then support the idea that $H_{c2}$ drops and goes to zero near the underdoped side of superconductor-insulator transition. Of course, both the coherence length and the upper critical field are different along in-plane and out-of-plane directions, but we assume that this anisotropy is temperature and doping independent. This might provide some evidence on the recent debate\cite{yayu,hc2} about how H$_{c2}$ behaves on underdoped cuprates. At the QCP the coherence length diverges for deeply underdoped cuprates and prevents BKT transition from occurring.

The case of ultrathin conventional NbN superconducting films is simpler. $\xi$ is isotropic and several tens of angstroms, to 2D behavior is always controlled by the thickness for ultrathin films:
\be
\label{3Dc2}
\xi \gtrsim d_{film}   ~~~(2D),
\ee
Therefore they are always in the 2D limit and BKT transition is always expected. Moreover, $H_{c2}$ measurements support the notion that $\xi$ increases when the QCP is approached,\cite{mondal_hc2} so that even at strong disorder films remain always in the 2D limit. Another way to think about the difference is, if we can reduce the thickness of cuprate films to a few unit cell so it is 2D by construction, then these films are similar to ultrathin conventional superconducting films and we should be able to recover the BKT downturn. That is what we indeed see in ultrathin Y$_{1-x}$Ca$_{x}$Ba$_{2}$Cu$_{3}$O$_{7-\delta}$ films.\cite{hetel} In this case, a BKT-like downturn is observed and it is robust down to the lowest level of doping.

\section{Conclusions}
Robust \BKT ~transitions, evident by sharp downturns of superfluid densities near T$_c$, are observed for ultrathin NbN films close to superconductor-insulator transition. They occur earlier than what 2-D XY model predicts and this is attributed to a relative small vortex-core energy. We observe that the vortex-core energy scales with the superconducting gap instead of the superfluid stiffness near the quantum critical point. Once included this effect, the BKT transition survives up to strong disorder, even though the sharp superfluid-density downturn observed in cleaner samples gets partly smeared out by the disorder-induced inhomogeneity of the system. The robustness of BKT transition is in direct contrast to similar studies on severely underdoped layered cuprates, which show no critical thermal fluctuations. This difference is attributed to the effect of a larger vortex-core energy or coherence length in deeply underdoped cuprates. Further investigation of both these mechanisms could shed new light on the nature of the superconductor-insulator transition in these unconventional superconductors.

\section{Acknowledgments}
Work at OSU was supported by DOE-Basic Energy Sciences through Grant No. FG02-08ER46533. L.B. acknowledges financial support by MIUR under FIRB2012(RBFR1236VV).


\begin{thebibliography}{cc}
\bibitem{nota_charged}
In a charged superfluid the occurrence of the BKT transition could be also limited by the screening effect of charged supercurrents, which cut-off the  logarithmic interaction between vortices at the Pearl length $L=2\lambda^2/d$. However, in underdoped cuprates the strong suppression of superfluid density due to correlations makes always the Pearl length larger than transverse sample dimensions, making this effect irrelevant.
\end{thebibliography}
\end{document}